\documentclass[prb,aps,twocolumn,superscriptaddress,showpacs]{revtex4}
\usepackage{amsfonts,amsmath,bm}

\newcommand{\st}{\tilde{\sigma}}

\usepackage{graphicx}

\begin{document}

\title{Photon-photon correlation statistics in the collective emission
  from ensembles of self-assembled quantum dots} 
\author{Fitria Miftasani}
\affiliation{Institute of Physics, Wroc{\l}aw University of
Technology, 50-370 Wroc{\l}aw, Poland}
\affiliation{Department of Physics, Bandung Institute of Technology,
  Ganesha 10 Bandung 40132, Indonesia} 
\author{Pawe{\l} Machnikowski}
 \email{Pawel.Machnikowski@pwr.edu.pl}  
\affiliation{Institute of Physics, Wroc{\l}aw University of
Technology, 50-370 Wroc{\l}aw, Poland}

\begin{abstract}
We present a theoretical analysis of the intensity autocorrelation for
the spontaneous emission from a planar ensemble of self-assembled
quantum dots. Using the
quantum jump approach, we numerically simulate the 
evolution of the system and construct photon-photon delay time
statistics that approximates the second order correlation function of
the field. The form of this correlation function in the
case of collective emission from a highly homogeneous ensemble qualitatively
differs form that characterizing an ensemble of independent emitters
(inhomogeneous ensemble of uncoupled dots). The signatures of
collective emission in the intensity correlations are observed also in
the case of an inhomogeneous but sufficiently strongly coupled
ensemble. Thus, we show that the second order correlation function of
the emitted field  provides a sensitive test of cooperative effects.
\end{abstract}


\maketitle

\section{Introduction} 
\label{sec:intro}

While the essential luminescence properties of quantum dots (QDs) are
determined by single-QD characteristics, the observed enhanced
luminescence of QD arrays and ensembles
\cite{scheibner07,mazur09,mazur10b} reveals physical effects beyond
this single emitter picture. Such an enhancement is attributed to collective
(cooperative) effects in emission, that is, the quantum-optical
phenomenon that leads to markedly non-exponential, peaked emission in
atomic samples \cite{skribanovitz73}. An obvious difference between
the QDs (``artificial atoms'') and the real, natural atoms is the
considerable inhomogeneity of optical transition energies in the
former case. With a natural emission line width of a single QD on the order
of a few $\mu$eV \cite{bayer02}, the typical ensemble distribution of
transition energies over several to a few tens of meV should preclude
any collective emission effects. Theoretical modeling reveals the
role of the interplay between this spectral inhomogeneity and the
coupling between the QDs \cite{sitek07a}. By assuming the presence of
some kind of electronic couplings (in addition to the fundamental but
weak dipole couplings) in the QD ensemble, the experimentally observed
enhancement of spontaneous emission in a planar ensemble
\cite{scheibner07} has been quantitatively reproduced
\cite{kozub12}. The proposed role of inter-QD couplings is also
consistent with the observed cooperative effects in QD chains
\cite{mazur10b}, where the presence of such couplings is much
more obvious \cite{mazur09}. 

The particular dependence of of the enhanced emission rate on
the ensemble size, spectral range, and excitation mode is consistent
with the concept of its collective nature and strongly supports this
interpretation \cite{scheibner07}. However, the observed time dependence
of the emission intensity for inhomogeneous ensembles of QDs remains
exponential \cite{scheibner07,mazur10b} (in contrast to the 
atomic superradiance) makes this effect purely quantitative and does
not allow one to exclude a formally possible conspiracy of different
factors that might simply shorten the exciton life time in the
QD samples used in the experimental studies. Therefore, one is
motivated to look for another characteristics 
of the emission signal, where the collective effects might be
manifested in a more qualitative way. Probably the most obvious option
is to look at the second-order correlation function of the emitted
radiation, which is experimentally accessible via detection of
photon-photon (intensity) correlations and is commonly applied to
characterize light 
fields emitted from QD systems \cite{ulrich07,Callsen2013,Davanco2014}. 
While the original approach, based on the
standard Hanbury-Brown and Twiss setup, limited the temporal
resolution of the measured correlation functions to the nanosecond
range, the recently developed experimental technique
\cite{Wiersig2009,Assmann2010a,Assmann2010b} gives access to second-order
correlations on much shorter time scales. In this method, one uses a streak
camera in the single photon counting mode to register the detection
traces of incident photons with picosecond resolution. This record of
time-labeled photodection events can than be used to determine the
statistical properties of a pulse light source with picosecond
resolution. Such a procedure has been successfully applied to characterize the light
field originating from a QD laser \cite{Wiersig2009,Assmann2010a}. 

Theoretical modeling of intensity correlations for QD ensembles focused
predominantly on cavity systems \cite{ulrich07}. General studies,
based on the Master equation approach, revealed strong photon bunching
in the emission form a few two-level emitters in cavities below the
lasing threshold \cite{temnov09} that oscillates with the number of
emitters \cite{auffeves11} and identified the role of dephasing
\cite{auffeves11}. For the description of QD 
systems, the model has been extended to account for multiple
excitation of a QD emitter and for the dephasing effects
\cite{ritter10}, as well as carrier-phonon interaction \cite{harsij12}
An extended theory has also been proposed \cite{ulrich07,su10},
based on the cluster expansion,
accounting for semiconductor-specific effects, like higher order
Coulomb and carrier-photon correlations or Pauli blocking. For
free-statnding systems, the intensity correlation has been modeled to
characterize two-photon emission from the biexciton cascade from a single
QD \cite{Callsen2013}.

In this paper, we present a theoretical study of second-order
correlations in the pulsed emission from a free-standing (not embedded
in a cavity) QD ensemble, using
the stochastic simulation (quantum jump) method. This work is motivated by 
the availability of the experimental technique mentioned above and
aims at identifying the signatures of cooperative effects in the
emission from such systems, in particular in photon-photon
correlations. We study also the dependence of the
correlation functions on the energy inhomogeneity and inter-dot
couplings in the ensemble. We find out that the form of the second order
correlation function in the case of emission from highly homogeneous
or sufficiently strongly coupled QD systems qualitatively differs from
that characterizing a field generated by an ensemble of independent
emitters. Therefore, the second order correlation function of
the emitted field  provides a sensitive test of cooperative effects in
the spontaneous emission.

The paper is organized as follows. Section \ref{sec:model-method} defines the
model and describes the stochastic simulation method used for the
numerical modeling. Sec.~\ref{sec:limits} is devoted to simple
limitiing cases that can be fully treated analytically. In
Sec.~\ref{sec:results}, we present and discuss 
the results of the simulations. Section~\ref{sec:concl} contains
general discussion concluding remarks.

\section{Model and simulation method} 
\label{sec:model-method}

In this section, we first define the model (Sec.~\ref{sec:model}), then describe the
stochastic method of quantum jumps used to simulate the dynamics
(Sec.~\ref{sec:method}) and, 
finally, define the estimates of the correlation functions in terms
of the photon emission events yielded by the stochastic simulation
(Sec.~\ref{sec:corr-fun}). 

\subsection{Model}
\label{sec:model}

The system to be modeled consists of  a planar, single-layer ensemble
of several
self-assembled QDs randomly and uniformly placed in the $xy$
plane, as in our earlier work \cite{kozub12}. 
The center-to-center distance between the QDs can not be lower than
10~nm (roughly the QD diameter).
The positions of the dots are denoted
by $\bm{r}_{\alpha}$, where $\alpha$ numbers the dots. 
Each QD 
is modeled as a point-wise two-level system (empty  
dot and one exciton) with the fundamental transition energy
$E_{\alpha}=\overline{E}+\epsilon_{\alpha}$, where $\overline{E}$ is
the average transition energy in the ensemble and $\epsilon_{\alpha}$
represent the energy inhomogeneity of the ensemble, described by a
Gaussian distribution with zero mean and standard deviation $\sigma$. 
We assume 
the dots to be coupled by an interaction $V_{\alpha\beta}$ which is composed
of long-range (LR) dipole interaction (dispersion force) and a
short-range (SR) coupling (exponentially decaying with the distance),
\begin{displaymath}
V_{\alpha\beta}=V^{(\mathrm{sr})}_{\alpha\beta}+V^{(\mathrm{lr})}_{\alpha\beta}.
\end{displaymath}
The long-range dipole coupling is described by 
\cite{stephen64,lehmberg70a,varfolomeev71,kozub12}
\begin{displaymath}
V^{(\mathrm{lr})}_{\alpha\beta}
=-\hbar\Gamma_{0}G(k_{0}r_{\alpha\beta}),\quad \alpha\neq\beta,
\end{displaymath}
and $V_{\alpha\alpha}=0$,
where $\bm{r}_{\alpha\beta}=\bm{r}_{\alpha}-\bm{r}_{\beta}$,
$\Gamma_{0}=
|d_{0}|^{2}k_{0}^{3}/(3\pi\varepsilon_{0}\varepsilon_{\mathrm{r}})$
is the spontaneous emission (radiative recombination) rate for a
single dot, $d_{0}$ is the magnitude of the interband dipole moment
(assumed identical for all the dots),
$\varepsilon_{0}$ is the vacuum permittivity,
$\varepsilon_{\mathrm{r}}$ is the relative dielectric constant of the
semiconductor, $k_{0}=n\overline{E}/(\hbar c)$,
$c$ is the speed of light,
$n=\sqrt{\varepsilon_{\mathrm{r}}}$ is the refractive index of the
semiconductor, 
and, for a heavy-hole transition in a planar ensemble,
\begin{displaymath}
G(x)  = -\frac{3}{8}\left(
\frac{\cos x}{x}+ \frac{\sin x}{x^{2}}+\frac{\cos x}{x^{3}}\right).
\end{displaymath}
The SR coupling is included in accordance with our previous work
\cite{kozub12}, which suggested its important role 
in the cooperative emission.
Only the overall magnitude and the finite range of this coupling are essentially
important, hence we model it by the simple exponential dependence
\begin{displaymath}
V^{(\mathrm{sr})}_{\alpha\beta}
=V_{0}e^{-r_{\alpha\beta}/r_{0}}.
\end{displaymath}

The density matrix then evolves according to the Master
equation\cite{lehmberg70a,kozub12}  
\begin{equation}
\label{evol}
\dot{\rho}=-\frac{i}{\hbar}[H_{0},\rho]+
\sum_{\alpha,\beta=1}^{N}\Gamma_{\alpha\beta}\left[ 
\sigma_{\alpha}\rho\sigma_{\beta}^{\dag}
-\frac{1}{2}\left\{ \sigma_{\beta}^{\dag}\sigma_{\alpha},\rho \right\}
 \right].
\end{equation}
Here the first term accounts for the unitary evolution of the ensemble
of coupled QDs with the Hamiltonian
\begin{equation}\label{H0}
H_{0} = \sum_{\alpha =1}^{N} \epsilon_\alpha \sigma_\alpha^\dagger
\sigma_\alpha 
+ \sum_{\alpha, \beta = 1}^N 
V_{\alpha \beta} \sigma_{\alpha}^{\dagger} \sigma_{\beta},
\end{equation}
where we introduce the transition
operators for the dots: the ``exciton annihilation'' operator
$\sigma_{\alpha}$, which annihilates an exciton in the dot $\alpha$,  
and the ``exciton creation'' operator $\sigma_{\alpha}^{\dagger}$
which creates an exciton in the dot $\alpha$
(the exciton number operator for the dot $\alpha$ is then
$\hat{n}_{\alpha}=\sigma_{\alpha}^{\dag}\sigma_{\alpha}$). 
In Eq.~\eqref{H0}, the first term describes the exciton energies in
the dots and the second one accounts for the inter-dot coupling. 
The second term in Eq.~\eqref{evol} describes the dissipation, that
is, the collective 
spontaneous emission process due to the coupling between the quantum
emitters (QDs) and their radiative environment (vacuum). 
Here $\Gamma_{\alpha\alpha}=\Gamma_{0},\quad
\Gamma_{\alpha\beta}=\Gamma_{\beta\alpha}=\Gamma_{0}F(k_{0}r_{\alpha\beta})$,
with 
\begin{eqnarray*}
F(x) & = & \frac{3}{4}\left(
\frac{\sin x}{x}-\frac{\cos x}{x^{2}}+\frac{\sin x}{x^{3}}
 \right),
\end{eqnarray*}
and $\{\ldots,\ldots\}$ denotes the anti-commutator. 

The simulations are performed for an ensemble of systems generated by
randomly placing a given number of 
QDs with a fixed surface density $\nu$ in the $xy$ plane and choosing
their fundamental transition energies from the Gaussian distribution. 
We assume the fully inverted  initial state corresponding to strong
excitation,
\begin{displaymath}
|\Psi_{0}\rangle = 
\prod_{\alpha=1}^{N}\sigma_{\alpha}^{\dag}|\mathrm{vac}\rangle,
\end{displaymath}
where $|\mathrm{vac}\rangle$ is the ``vacuum'' state, that is,
the crystal ground state state with filled valence band states and
empty conduction band states (no excitons in the QDs). Such a state
can be created by strong optical excitation. Exciton injection to QD
ensembles can also be controlled by an external electric field
\cite{Baumgartner2010}.

\subsection{Stochastic simulation method}
\label{sec:method}

We model the evolution of the spontaneous emission process numerically,
using the stochastic simulation approach  (quantum
jump method) \cite{Breuer,Gardiner2004}, which considerably reduces
the computational load and 
allows us to study larger QD ensembles than those tractable by a direct
integration of the Master equation. The starting point is 
Eq.~(\ref{evol}) written in the equivalent form
\begin{equation}
\label{evol-diag}
\dot{\rho}=
-\frac{i}{\hbar}(H_{\mathrm{eff}}\rho-\rho H_{\mathrm{eff}}^{\dag})+
\sum_{i=1}^{N}\tilde{\Gamma}_{i}
\st_{i}\rho\st_{i}^{\dag},
\end{equation}
where 
\begin{displaymath}
H_{\mathrm{eff}}=H_{0} +\frac{\hbar}{2i}\sum_{\alpha\beta}
\Gamma_{\alpha\beta} \sigma_{\beta}^{\dag}\sigma_{\alpha},
\end{displaymath}
$\tilde{\Gamma}_{i}$ are the eigenvalues of the
positive-definite, symmetric, real matrix $\Gamma_{\alpha\beta}$
obtained via diagonalization of the latter with the unitary matrix
$u_{i\alpha}$,
\begin{displaymath}
\sum_{\alpha,\beta=1}^{N} u_{i\alpha}\Gamma_{\alpha\beta}u^{*}_{j\beta}
=\tilde{\Gamma}_{i}\delta_{ij},
\end{displaymath}
and 
\begin{displaymath}
\st_{i}=\sum_{\alpha=1}^{N}u_{j\alpha}^{*}\sigma_{\alpha}.
\end{displaymath}
The Master equation (\ref{evol}) can then be equivalently
replaced by a 
stochastic simulation  \cite{Breuer,Gardiner2004} in which the
unnormalized state vector 
$|\Psi\rangle$ evolves along continuous trajectories in the Hilbert
space according to the equation of motion
\begin{equation}\label{cont}
i\hbar\frac{d}{dt}|\Psi\rangle = H_{\mathrm{eff}}|\Psi\rangle
\end{equation}
interrupted by one of the discontinuous jumps
\begin{equation}\label{jump}
|\Psi\rangle \to 
\frac{\sqrt{\tilde{\Gamma}_{i}}\st_{i}|\Psi\rangle}{
|\sqrt{\tilde{\Gamma}_{i}}\st_{i}|\Psi\rangle|^{1/2}},\quad
i=1,\ldots,N.
\end{equation}
The duration of each continuous evolution period (the waiting time for
the next jump) is a randorm variable with the cumulative distribution
function 
\begin{equation}\label{CDF}
F(t)=1-\langle \Psi(t) | \Psi(t)\rangle.
\end{equation}
The relative
probability that the $i$th jump will take place is 
\begin{equation}\label{rel-probab}
p_{i}=\frac{
\tilde{\Gamma_{i}}\langle \Psi(t) |\st_{i}^{\dag}\st_{i}|
  \Psi(t)\rangle}{
\sum_{i=1}^{N}\tilde{\Gamma_{i}}\langle \Psi(t) |\st_{i}^{\dag}\st_{i}|
  \Psi(t)\rangle}.
\end{equation}
This prescription is implemented by evolving the system after each
jump according to Eq.~(\ref{cont}) until $\langle \Psi(t) |
\Psi(t)\rangle=X$, where $X$ is a random variable uniformly
distributed on $[0,1)$, obtained from the quasi-random number
generator of a computer, and then selecting the jump at random
according to Eq.~(\ref{rel-probab}).  

\subsection{Correlation functions}
\label{sec:corr-fun}

Each jump can be identified with a photon emission. Hence, 
the
statistics of the jump times over many repetitions of the numerical
experiment can be used to approximate the luminescence intensity. 
To this end, we model the repeated pulsed excitation of the system by
periodically setting the initial condition and simulating the
subsequent emission. In each repetition, 
we register 
the time stamp for each jump and build the discrete statistics using a
time bin $\Delta t$,
\begin{displaymath}
I(t)\approx \frac{\langle N(t) \rangle}{N_{\mathrm{rep}}\Delta t},
\end{displaymath}
where $N(t)$ is the number of photons emitted in the time interval
$\Delta t$ containing $t$, $N_{\mathrm{rep}}$ is the number of  repetitions of the
simulation, and $\langle\ldots\rangle$ denotes averaging over the
repetitions.
We use the same method to
compute the second order correlation function $G^{(2)}(t_{1},t_{2})$ of the
emitted radiation,
\begin{displaymath}
\mathcal{G}^{(2)}(t_{1},t_{2}) \approx
\frac{\langle N(t_{1})N(t_{2})\rangle}{N_{\mathrm{rep}}(\Delta t)^{2}}.
\end{displaymath}
For our discussion, we introduce a pre-factor, inversely
proportional to the number of photon pairs in a single pulse, which
assures that results obtained
for ensembles of different sizes are comparable. Thus, we define
\begin{displaymath}
G^{(2)}(t_{1},t_{2}) \approx 
\frac{2}{N(N-1)} \mathcal{G}^{(2)}(t_{1},t_{2}),
\end{displaymath}
where $N$ is the number of QDs in the ensemble. 
Note that, apart from
this uniform re-scaling, the function $G^{2}$ has the form and meaning
of an unnormalized correlation function. The standard way to
normalize the correlations is to use the product of intensities at the
two times involved,
\begin{displaymath}
g^{(2)}(t_{1},t_{2})=\frac{G^{(2)}(t_{1},t_{2})}{I(t_{1})I(t_{2})}.
\end{displaymath}
 Due to the
normalization, the two 
functions $g^{(2)}(t_{1},t_{2})$ and $g^{(2)}(\tau)$ represent the
intrinsic properties of the field, irrespective of the photon
collection and detection efficiency.

Although the two-time correlation function is necessary to provide the full
information on second order correlations in the non-stationary
evolution, a simpler quantitative characteristics might be useful to
characterize the collective emission. 
Here, we will use the integrated correlation function that
depends only on $\tau=t_{2}-t_{1}$ and represents the statistics of
the delay times $\tau$ between pairs of emission events (as used in
the experimental studies \cite{Wiersig2009,Assmann2010b}),
\begin{displaymath}
G^{(2)}(\tau)=\int_{0}^{T_{\mathrm{rep}/2}} dt G^{(2)}(t,t+\tau)
\end{displaymath}
and the corresponding normalized function
\begin{equation}\label{g2-norm}
g^{(2)}(\tau)=\frac{\int_{0}^{T_{\mathrm{rep}/2}}
  dt \mathcal{G}^{(2)}(t,t+\tau)}{%
\int_{0}^{T_{\mathrm{rep}/2}}  I(t)I(t+\tau)}.
\end{equation}
Eq.~\eqref{g2-norm} can be viewed as an average of the normalized
function $g^{(2)}(t_{1},t_{2})$ weighted by the product of relative (normalized)
intensities at the two times \cite{Assmann2010b}.

This stochastic procedure outlined above is equivalent to the standard approach
based on the Master equation and quantum regression theorem in the
sense of the statistics of an arbitrary sequence of photodetection measurements
\cite{Gardiner2004}. Moreover, it exactly follows the experimental
procedure used to calculate the correlation functions from a
time-resolved sequence of photodetection events
\cite{Wiersig2009,Assmann2010b}. 
The advantage of applying the stochastic scheme is not only reducing
the description from the density matrix ($\sim N^{2}$ variables) to the
state vector ($\sim N$ variables) level. An additional benefit follows
from the fact that each jump described in Eq.~(\ref{jump}) reduces the
average number of excitons in the ensemble exactly by one. Since coherences
between states with a different number of excitons, if initially
absent, cannot appear as a result of the evolution, this allows us to
integrate the equation of motion within a subspace with a given number
of excitons, which further considerably reduces the size of the numerical problem.

In our simulations, we  use the parameters :
$\Gamma_{0}=2$~ns$^{-1}$, $n=2.6$, the average transition energy of
the QD ensemble $\overline{E}=2.59$~eV and the QD surface density 
$\nu=10^{11}$ /cm$^{-2}$. For the
tunnel coupling we choose  the range $r_{0}=15$~nm, while its
amplitude $V_{0}$ is used as a parameter (the dipole coupling is
always present, unless explicitly noted). The repetition period
$T_{\mathrm{rep}}=12\Gamma^{-1}=6$~ns is sufficient to avoid
noticeable overlap with the tail from the previous repetition. We use
the time bin $\Delta t=0.06$~ns and perform $N_{\mathrm{rep}}=10^{6}$
repetitions for 2 QDs, $N_{\mathrm{rep}}=10^{5}$
repetitions for 6 and 10 QDs (unless noted otherwise), and
$N_{\mathrm{rep}}=10^{4}$ repetitions for 16 QDs.

\section{Special limits}
\label{sec:limits}

In order to set some frame for the discussion of numerical results to
be presented in Sec.~\ref{sec:results}, 
in this section we discuss easily obtainable
analytical results pertaining to the limits of independent emitters
and to fully collective emission from two QDs.

For a single QD, the probability of photon emission in the time
interval $(t,t+dt)$ is $f_{1}(t)=\Gamma e^{-\Gamma t}dt$. 
Within our model, $N$ QDs emit $N$ photons. 
If the emission from each
QD is independent (uncorrelated) then the joint probability
density for
detecting the (distinguishable) photons at times $t_{1},\ldots,t_{N}$ is 
$f(t_{1},\ldots,t_{N})=f_{1}(t_{1}\ldots f_{1}(t_{N}))$, where $t_{i}$
is the time of photon emission from the $i$th QD. The
two-photon correlation function is the probability density for
detecting a pair of photons (emitted by whichever pair of QDs) around
times $t$ and $t'$, 
\begin{align*}
\mathcal{G}^{(2)}(t,t') &=N(N-1)\int dt_{3} \ldots \int dt_{N}
f(t,t',t_{3}\ldots,t_{N}) \\
& =N(N-1)\Gamma^{2} e^{-\Gamma (t+t')},
\end{align*}
where the combinatorial pre-factor results from adding identical
expressions for each pair of QDs. The intensity in the independent
emission case is $I(t)=N\Gamma e^{-\Gamma t}$. Hence, according to
Eq.~\eqref{g2-norm}, 
\begin{displaymath}
g^{(2)}(\tau)=1-\frac{1}{N},
\end{displaymath}
which is the same expresson as for a stationary $N$-photon field.

In the case of two identical, uncoupled QDs in the Dicke limit of
vanishing inter-dot distance (hence $\Gamma_{12}=\Gamma_{0}$), the
photon statistics might be found from the solution to the
Master equation \eqref{evol}, using the quantum regression theorem. 
Alternatively, one can resort to the quantum jump picture: From
Eq.~\eqref{cont} one finds in this case for the evolution of the
unnormalized biexciton (fully inverted) state 
$\langle \Psi(t)|\Psi(t)\rangle=e^{-2\Gamma t}$,
from which, according to Eq.~\eqref{CDF}, the probability density for the first
jump time is $f_{1\mathrm{st}}(t)=2\Gamma e^{-2\Gamma t}$. According
to Eq.~\eqref{jump}, upon the first emission the system is projected
on the single-exciton Dicke state (that is, the fully symmetric
superposition of states with a given exciton number, here single
exciton states). Using again Eq.~\eqref{cont},
one finds the evolution of the corresponding density matrix, 
$\langle \Psi(t+\tau)|\Psi(t+\tau)\rangle=e^{-2\Gamma \tau}$, 
where $\tau$ is the time
interval after the first emission. Hence, the probability density
for the second emission event, conditioned on the first emission at
time $t$, is $f_{2\mathrm{nd}}(\tau|t)=2\Gamma e^{-2\Gamma \tau}$.
Note that this conditional probability density does not depend
on the first emission time but only on the time delay. The joint
probability, that is, the unnormalized 2nd order correlation function
is then $\mathcal{G}^{(2)}(t,t+\tau)= 
f_{1\mathrm{st}}(t)f_{2\mathrm{nd}}(\tau |t) =
4\Gamma^{2} e^{-2\Gamma (t+\tau)}$, $\tau>0$,
or, 
\begin{displaymath}
\mathcal{G}^{(2)}(t,t')= 4\Gamma^{2} e^{-2\Gamma \max(t,t')}
\end{displaymath}
and depends only on the second (later) emission time.
The intensity is calculated as the total probability of emitting the
first or the second photon at a given time
\begin{align*}
I(t) &=
\int_{0}^{\infty}d\tau\mathcal{G}^{(2)}(t,t+\tau)
+\int_{0}^{t}dt' \mathcal{G}^{(2)}(t',t) \\
& = 2\Gamma(1+2\Gamma t)e^{-2\Gamma t}.
\end{align*}
From this, according to Eq.~\eqref{g2-norm}, one finds
\begin{displaymath}
g^{(2)}(\tau)=\frac{4}{6\Gamma \tau +5},
\end{displaymath}
which shows a power-law decay and therefore qualitatively differs from
the constant intensity correlation function for independent emitters.

\section{Results and discussion}
\label{sec:results}

In this section we present and discuss the results of our
simulations. First, in Sec. \ref{sec:results-int-correl}, we analyze
the time-resolved intensity of the 
emitted radiation and the full, two-time correlation function. Next,
in Sec.~\ref{sec:delay}, we focus on the integrated correlation
function representing the statistics of delay times between two photon
emission events.

\subsection{Intensity and correlations}
\label{sec:results-int-correl}

\begin{figure}[tb]
\begin{center}
\includegraphics[width=85mm]{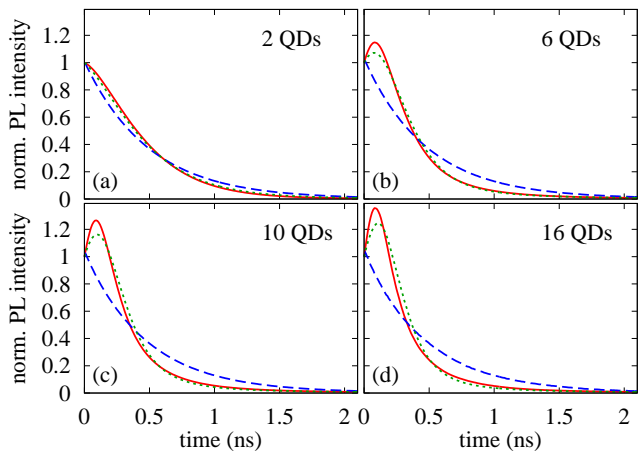}
\end{center}
\caption{\label{fig:intensity}The intensity of luminescence from an
  ensemble of 2,6,10, and 16 QDs. Red solid lines: $\sigma=0$,
  $V_{0}=0$; blue dashed lines:  $\sigma=0.1$~meV, $V_{0}=0$; green dotted 
  lines:  $\sigma=0.1$~meV, $V_{0}=0.5$~meV.}
\end{figure}

Fig.~\ref{fig:intensity} shows the PL intensity as a function of time for
the emission from ensembles of several QDs. If the QDs are identical,
$\sigma=0$ 
and coupled only by the weak dipole interactions, $V_{0}=0$ (red solid lines), the
luminescence decay is non-exponential and develops a non-monotonicity
(a superradiant peak) as the number of QDs grows. This effect is
completely destroyed already by a weak inhomogeneity of the transition energies in
the ensemble, $\sigma=0.1$~meV, well below the degree of inhomogeneity
expected in a 
real sample (blue dashed lines). The collective emission effect, with
the peaked luminescence, is restored if the QDs are sufficiently
strongly coupled, $\sigma=0.1$~meV, $V_{0}=0.5$~meV (green dotted
lines) \cite{sitek07a}. Here, we see 
the first difference between the case of 2 QDs and a larger number of
QDs: the evolution of a strongly coupled 2-QD system nearly exactly
follows that for identical dots, while in the other cases, the
original photoluminescence decay curve is not completely restored. The
reason is that in the 2 QD case the single-exciton eigenstate of a
strongly coupled system coincides with the Dicke state, while in a
larger ensemble of randomly distributed dots the Dicke 
states for various exciton numbers are typically not exact eigenstates
of the system, although they 
may have an enhanced overlap with them. 

\begin{figure}[tb]
\begin{center}
\includegraphics[width=85mm]{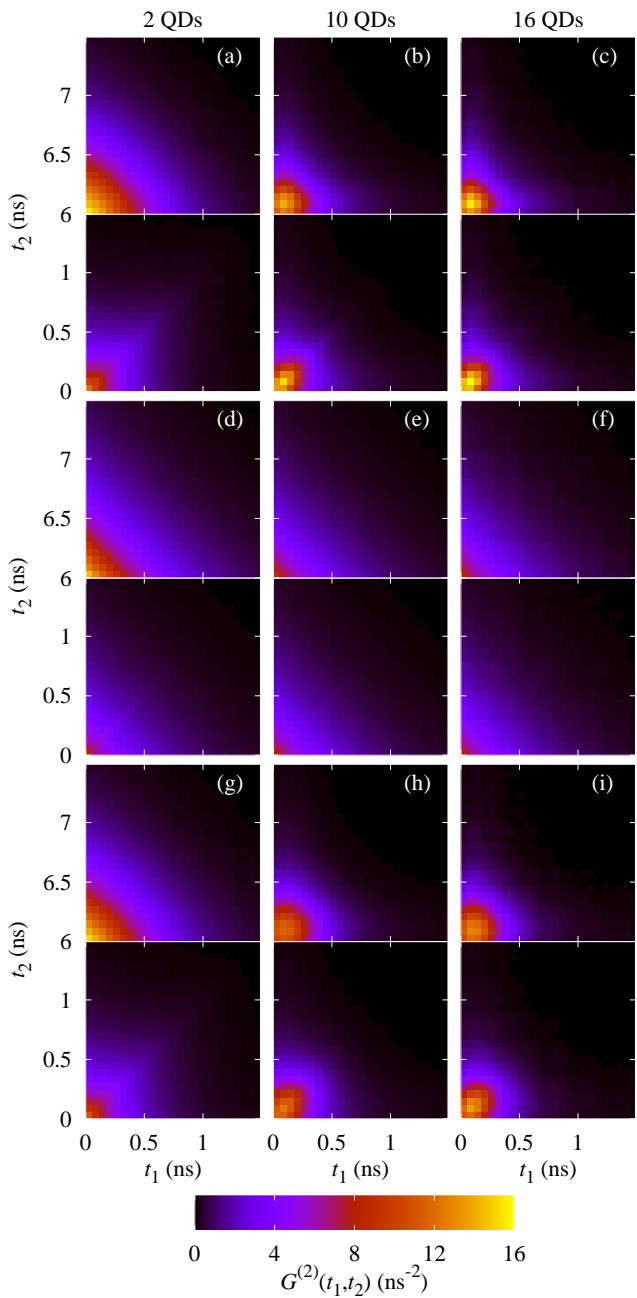}
\end{center}
\caption{\label{fig:maps}The two-time correlation function
  $g^{(2)}(t_{1},t_{2})$ for 2,10, and 16 QDs. (a)-(c): $\sigma=0$,
  $V_{0}=0$; (d)-(f): 
  $\sigma=0.1$~meV, $V_{0}=0$ (g)-(i):  $\sigma=0.1$~meV,
  $V_{0}=0$. The range of $t_{2}$ covers two repetition periods with a
  certain time interval, where the signal is nearly null, removed.}
\end{figure}

The interplay between the inhomogeneity and coupling is reflected also
in the two-time second order correlation of the photon emission
events, as shown in Fig.~\ref{fig:maps}. Each correlation map is shown
over one repetition period of the numerical experiment in $t_{1}$ and
over two repetition periods in $t_{2}$ (the correlation function for
any higher period is the same). The maps show clear
differences between the correlations within one repetition period and
between different repetition periods, as well as between the cases
characterized by different inhomogeneity and inter-dot coupling. 

Figs.~\ref{fig:maps}(a)-(c) present the correlation functions for
ensembles of identical dots coupled only by the fundamental, weak
dipole interactions. 
The photon emission events in different repetitions of the experiment
are uncorrelated. Hence, if $0<t_{1}<T_{\mathrm{rep}}$ and 
$T_{\mathrm{rep}}<t_{2}<2T_{\mathrm{rep}}$ then 
$g^{(2)}(t_{1},t_{2})\sim I(t_{1})I(t_{2})$, where $I(t)$ is the
luminescence intensity. For identical dots, the emission is
non-exponential, with a peak developing at $t>0$ if the number of dots
exceeds 2. This is simply reflected in the shape of the correlation
function, which has a maximum at $t_{1}=t_{2}>0$ for
$N_{\mathrm{QD}}>2$. The picture is very different for the
correlations between events within the same repetition period, when
the photons are emitted in the course of a single instance of the
system evolution. Here, again, the case of 2 QDs (Fig.~\ref{fig:maps}(a))
is exceptional: As explained in Sec.~\ref{sec:limits}, after 
one photon is emitted, there is exactly one left and, according to
Eq.~\eqref{jump}, the system is
projected always on the same single-exciton Dicke
state. Therefore, the system evolution after the first emission does
not depend on the first emission time. As a result, the correlation function depends
only on $\max(t_{1},t_{2})$, hence the characteristic square shaped
form of $G^{(2)}(t_{1},t_{2})$ visible in
the lower panel of Fig.~\ref{fig:maps}(a). For a larger number of QDs, the
system is likely to contain more exciton after emission events that
happened at earlier times, hence the further evolution and, in
consequence, the correlation function, depends on both time
arguments. The square shaped profile of $G^{(2)}$ still remains to
some extent visible for 6
QDs (Fig.~\ref{fig:maps}(b)) but is much less pronounced already for 16
QDs (Fig.~\ref{fig:maps}(c)).

If the dispersion of transition energies becomes large enough to
prevent the collective emission, as in the cases shown in
Figs.~\ref{fig:maps}(d)-(f), then the correlation function shows none of
the features discussed above. Now, each photon is emitted by a single
and independently evolving QD, hence different emission events are
always independent and the correlation function becomes proportional
to a product of two exponentially decaying luminescence intensities,
$g^{(2)}(t_{1},t_{2})\sim \exp[-\Gamma(t_{1}+t_{2})]$ (see
Sec.~\ref{sec:limits}). This dependence 
on $t_{1}+t_{2}$ only is clear in Figs.~\ref{fig:maps}(d)-(f), where the
only difference between the correlations within and between the
repetitions results from the combinatorial factor ($N(N-1)$
vs. $N^{2}$ photon pairs).

As discussed above, sufficiently strong coupling can restore the
characteristic features related to collective luminescence. The
shape of the correlation function is also to some extent restored
(Figs.~\ref{fig:maps}(g)-(i)) and becomes similar (although not quite
identical) to that observed for identical dots.

From now on, we will focus on the correlations within one repetition
period. 

\subsection{Delay time statistics}
\label{sec:delay}

\begin{figure}[tb]
\begin{center}
\includegraphics[width=85mm]{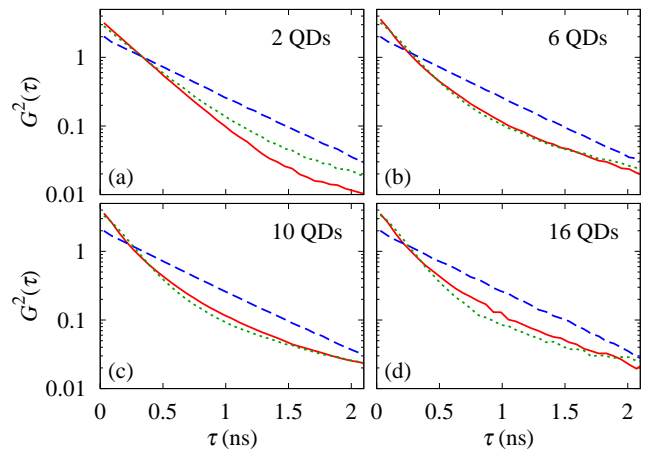}
\end{center}
\caption{\label{fig:g2}The unnormalized integrated correlation function $G^{(2)}(\tau)$
  as a 
  function of the delay $\tau$ for the emission from an
  ensemble of 2,6,10, and 16 QDs. 
 Red solid lines: $\sigma=0$,
  $V_{0}=0$; blue dashed lines:  $\sigma=0.1$~meV, $V_{0}=0$; green dotted 
  lines:  $\sigma=0.1$~meV, $V_{0}=0.5$~meV.}
\end{figure}

A simple but still very informative characteristics of intensity
correlations is the delay-dependent correlation function
$G^{(2)}(\tau)$, defined in Sec.~\ref{sec:corr-fun},
that reflects the statistics of the delay times between pairs of
emission events averaged over all the evolution time. This correlation
function is shown in Fig.~\ref{fig:g2} for ensembles of 2,6,10, and 16
QDs. This function is very close to exponential for uncorrelated decay in a strongly
inhomogeneous system (blue dashed lines) but becomes non-exponential
for a system of more than two identical QDs that emit cooperatively (red
lines). This non-exponential dependence on the delay time is restored
also by sufficiently strong interactions (green dotted lines).

\begin{figure}[tb]
\begin{center}
\includegraphics[width=85mm]{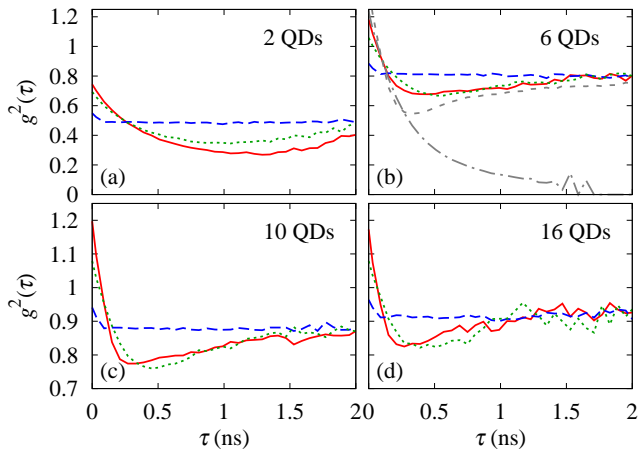}
\end{center}
\caption{\label{fig:g2-norm}The normalized integrated correlation
  function $g^{(2)}(\tau)$   as a 
  function of the delay $\tau$ for the emission from an
  ensemble of 2,6,10, and 16 QDs.  Red solid lines: $\sigma=0$,
  $V_{0}=0$; blue dashed lines:  $\sigma=0.1$~meV, $V_{0}=0$; green dotted 
  lines:  $\sigma=0.1$~meV, $V_{0}=0.5$~meV. In (c), the grey dashed
  and dash-dotted  lines show the results in the Dicke limit and in
  the case of vanishing dipole coupling, respectively ($10^{7}$
  repetitions in the latter case).} 
\end{figure}

Since the function $G^{(2)}(\tau)$ depends on the actual number of
detection events, its magnitude reflects the experimental conditions
and only its qualitative features are meaningful. In contrast, its
normalized counterpart, $g^{(2)}(\tau)$, defined by
Eq.~\eqref{g2-norm}, yields intrinsic quantitative 
information on the photon correlations. This function is shown in
Fig.~\ref{fig:g2-norm}. As a result of the
normalization by field intensities
the form of this function differs considerably from the unnormalized
one.  

In the case of collective emission (homogeneous ensembles or
a sufficiently strong coupling, red solid and green dotted lines in
Fig~\ref{fig:g2-norm}, respectively), the value of $g^{(2)}(0)$ is
below 1 for a small number of emitters, indicating a non-classical
nature of the field (Fig.~\ref{fig:g2-norm}(a)), while it shows a
bunching effect, $g^{(2)}(0)>1$, 
for a larger number of QDs (Fig.~\ref{fig:g2-norm}(b-d)).
At finite delays, the normalized correlation function in this case
drops down to reach 
a minimum at a certain time on the order of the spontaneous emission
time longer for smaller ensembles) and then increases again up to the
a value of $(N-1)/N$ which characterizes uncorrelated emission. The
fact that this holds also for a perfectly homogeneous system coupled
only by the very weak dipole interactions is quite striking. For a
strictly superradiant decay, the correlation function should decay to
0, as discussed for the special case of 2 QDs in
Sec.~\ref{sec:limits}. Our system differs from that formal limiting
case in two respects:
First, the distances between the dots are finite (although small),
which affects the 
collective nature of the coupling (technically,
$\Gamma_{\alpha\beta}<\Gamma_{0}$ for $\alpha\neq\beta$ in
Eq.~\eqref{evol}). Second, the weak but non-zero dipole coupling is
always present, as it is
of fundamental nature and inseparable from the spontaneous emission
process. In order to assess the role of these two factors, in
Fig.~\ref{fig:g2-norm}(b) we have included the results for two kinds hypothetical
systems: the Dicke limit (negligible distance between the emitters,
hence $\Gamma_{\alpha\beta}=\Gamma_{0}$ for all $\alpha,\beta$) with
the dipole interactions as for the actual inter-dot distances in our
ensemble, as well as an ensemble without any couplings (grey short-dashed and dash-dotted
lines, respectively). The comparison of these formal results shows
that even in the Dicke limit the behavior of the intensity correlation
is similar to the actual one and only upon removing the coupling the
non-monotonic behavior of $g^{(2)}(\tau)$ is turned into a monotonic
decay. Hence, we conclude that the fundamental
dipole coupling, which is too weak to induce any observable traces of
collective emission in the luminescence intensity from an
inhomogeneous ensemble \cite{kozub12},
qualitatively changes the form of the normalized second-order
correlation function. 

In the case of an inhomogeneous and weakly coupled system, when the
decay of both the intensity and the correlations is indistinguishable
from exponential (blue dashed lines), the value of the correlation function is
almost everywhere constant and equal to $(N-1)/N$, as for uncorrelated
emitters. An exception are very short delays, where it shows a weak
but clear
increase towards $\tau=0$, indicating that some small degree of
cooperativity is present in the emission. Hence, also in this case the correlations
are more sensitive to the system properties than the intensities.

\begin{figure}[tb]
\begin{center}
\includegraphics[width=85mm]{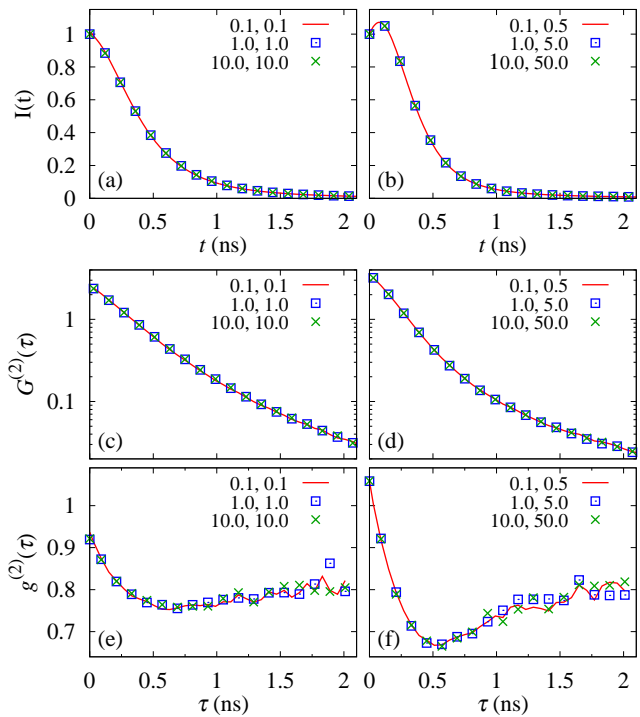}
\end{center}
\caption{\label{fig:equiv}Comparison of luminescence intensities and
  second-order correlation functions for ensembles with different
  magnitudes but a fixed ratio between the essential parameters:
  energy inhomogeneity and coupling strength. The numbers describing
  the lines and symbols indicate the values of $\sigma$ and $V_{0}$ in
  meV.} 
\end{figure}
   
So far, we have discussed ensembles with rather low energy
inhomogeneities. However, one can expect that, as soon as both the
energy inhomogeneity and typical coupling strengths are much larger
than the natural emission line width, the ensemble luminescence should
only depend (up to trivial scaling of the time axis) on the ratio of
these two parameters. This is strictly the case for 2 QDs,
where the energy difference and coupling strength fully characterize
the system \cite{sitek07a}. Although for larger ensembles both these
parameters merely characterize a distribution of random values (which
is, in addition, non-trivial in the case of the distance-dependent
couplings), 
Fig.~\ref{fig:equiv} shows that such a universal scaling 
is perfectly valid over two orders of magnitude of these parameters
also in the case of larger QD ensembles (here 6 QDs). This holds true
not only for the luminescence intensity (Fig.~\ref{fig:equiv}(a,b))
but also for the unnormalized and normalized intensity correlation
functions $G^{(2)}(\tau)$ and $g^{(2)}(\tau)$ (Fig.~\ref{fig:equiv}(c-f)).

\begin{figure}[tb]
\begin{center}
\includegraphics[width=85mm]{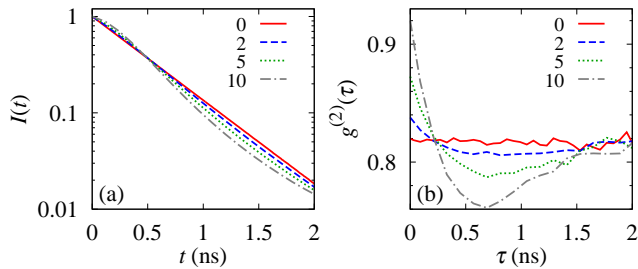}
\end{center}
\caption{\label{fig:I-g2} Comparison of the decay of the luminescence
  intensity $I(t)$ (a) and  the normalized correlation function
  $g^{(2)}(\tau)$ (b) for 6 QDs with $\sigma=10$~meV and various
  strengths of the coupling $V_{0}$ as shown (in meV). Here, the
  results have been averaged over $10^{6}$ repetitions.} 
\end{figure}

As the the cooperative nature of the emission proces manifests itself
in a qualitative way in the intensity correlations, the correlations
statistics may be a better experimental test for such collective
effects. Indeed, as we show in Fig.~\ref{fig:I-g2}, the non-monotonic
form of the correlation function $g^{(2)}(\tau)$ [Fig.~\ref{fig:I-g2}(a)] appears already for
inhomogoneous systems in which the the coupling is so weak that the
deviation from the exponential luminescence decay [Fig.~\ref{fig:I-g2}(b)] is almost
unnoticeable (in particular, blue dashed lines in Fig.~\ref{fig:I-g2},
corresponding to $V_{0}=2$~meV$=0.2\sigma$). 

\begin{figure}[tb]
\begin{center}
\includegraphics[width=85mm]{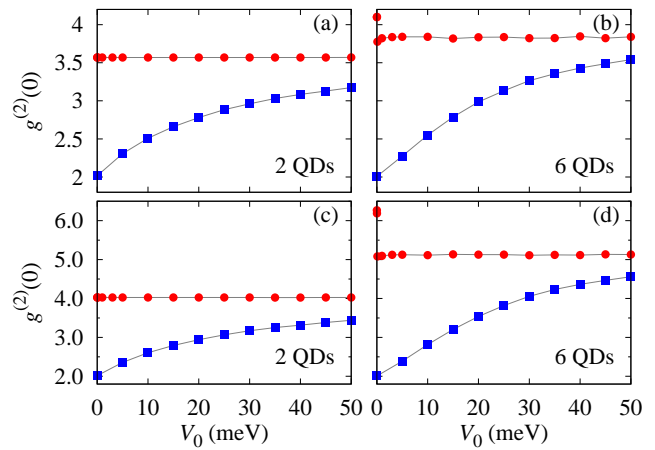}
\end{center}
\caption{\label{fig:g20}The correlation function $g^{(2)}(0)$ as a
  function of the coupling parameter $V_{0}$ for a realistic ensemble
  of two and six QDs  (a,c) and assuming the Dicke limit of vanishing
  geometrical size of the ensemble (b,d). Red circles: identical QDs; Blue squares:
  non-identical QDs with $\sigma=10$~meV.} 
\end{figure}

A single parameter, quantitatively characterizing the photon-photon
correlations, is the value of the correlation function at zero delay,
$g^{(2)}(0)$. This is shown in Fig.~\ref{fig:g20}(a,b) for ensembles of 2 and 6
identical QDs (red circles) and for inhomogeneous QD ensembles (blue
squares). The effect of the coupling is different in the two
cases. An ensemble of identical emitters emits superradiantly in the
absence of coupling, while the coupling partly destroys the cooperative
emission effect by lifting the degeneracy of the states with a given
number of excitons and thus perturbing the Dicke states, which are no
more energy eigenstates. As 
a result, the value of $g^{(2)}(0)$ drops down. An
exception is the case of 2 QDs, where the eigenstate of coupled,
identical dots coincides with the Dicke state, as discussed
above. On the contrary, if the system is energetically inhomogeneous
(blue squares), the major effect of the coupling is to overcome the
inhomogeneity and to form partly symmetrized eigenstates, which,
although not completely identical with the Dicke state for
$N_{\mathrm{QD}}>2$, restore the cooperative emission and lead to an
increase of  $g^{(2)}(0)$. 

We note at this point that, apart from the
interplay of inhomogeneity and coupling, the cooperative emission form
a QD ensemble is affected also by the dependence of the off-diagonal
transition rates $\Gamma_{\alpha\beta}$ on the inter-dot
distance. Although the spatial sizes of the ensembles modeled here are
always smaller than the wave length (about 1~$\mu$m in the medium,
compared to 30~nm average distance between the neighboring QDs), the
finite-size effect is quantitatively noticeable. This is illustrated
by Fig.~\ref{fig:g20}(c,d), where the results as in the two upper panels
are shown but this time assuming the (formal) Dicke limit of a very small
ensemble, with $\Gamma_{\alpha\beta}=\Gamma_{0}$. While the
qualitative dependence of $g^{(2)}(0)$ is the same as in the realistic
case, the values are clearly higher.

\section{Conclusions}
\label{sec:concl}

We have studied the second order correlation function (intensity
autocorrelation) on sub-nanosecond time scales for the spontaneous
emission from inhomogeneous 
planar ensembles of coupled QDs using the quantum jump approach to
numerical simulations of the open system dynamics, which mimics a
recently developed experimental technique. 

We have shown that the cooperative nature of the emission manifests
itself not only by non-exponential decay of luminescence, but also in
the particular form of both the normalized and unnormalized
correlation functions.  The collective emission case corresponds either to the
rather hypothetical limit of a very homogeneous QD ensemble or to a
realistic case of en inhomogeneous ensemble of coupled dots. 
The signatures of collective emission in the intensity correlations
are the same in both these cases.  

The normalized correlation function is of particular interest as it
shows qualitative features (non-monotonic delay dependence) for weakly
coupled inhomogeneous systems for which the non-exponential character of
luminescence decay is so weak that it may be hard to notice
experimentally. Thus, we conclude that the photon-photon delay time
statistics may be a better experimental probe of cooperative effects
than the time-resolved luminescence intensity.

\acknowledgments
This work was supported in part by the
Polish National Science Centre (Grant No.~2011/01/B/ST3/02415). 
FM acknowledges support from the Erasmus Mundus StrongTies Program of
the UE. 
Calculations have been partly carried out in Wroclaw Centre for
Networking and Supercomputing (http://www.wcss.wroc.pl), Grant No. 203.


\begin{thebibliography}{10}

\bibitem{scheibner07}
M. Scheibner, T. Schmidt, L. Worschech, A. Forchel, G. Bacher, T. Passow, and
  D. Hommel, Nat. Phys. {\bf 3},  106  (2007).

\bibitem{mazur09}
Y.~I. Mazur, V.~G. Dorogan, J. {E. Marega}, G.~G. Tarasov, D.~F. Cesar, V.
  Lopez-Richard, G.~E. Marques, and G.~J. Salamo, Appl. Phys. Lett. {\bf 94},
  123112  (2009).

\bibitem{mazur10b}
Y. Mazur, V.~G. Dorogan, E. Marega, D.~F. Cesar, V. Lopez-Richard, G.~E.
  Marques, Z. Zhuchenko, G.~G. Tarasov, and G.~J. Salamo, Nano. Res. Lett. {\bf
  5},  991  (2010).

\bibitem{skribanovitz73}
N. Skribanowitz, I.~P. Herman, J.~C. MacGilvray, and M.~S. Feld, Phys. Rev.
  Lett. {\bf 30},  309  (1973).

\bibitem{bayer02}
M. Bayer and A. Forchel, Phys. Rev. B {\bf 65},  041308  (2002).

\bibitem{sitek07a}
A. Sitek and P. Machnikowski, Phys. Rev. B {\bf 75},  035328  (2007).

\bibitem{kozub12}
M. Kozub, b. Pawicki, P. Machnikowski, and Å. Pawicki, Phys. Rev. B {\bf 86},
  121305  (2012).

\bibitem{ulrich07}
S.~M. Ulrich, C. Gies, S. Ates, J. Wiersig, S. Reitzenstein, C. Hofmann, A.
  L\"{o}ffler, A. Forchel, F. Jahnke, and P. Michler, Phys. Rev. Lett. {\bf
  98},  043906  (2007).

\bibitem{Callsen2013}
G. Callsen, A. Carmele, G. H\"{o}nig, C. Kindel, J. Brunnmeier, M.~R. Wagner,
  E. Stock, J.~S. Reparaz, A. Schliwa, S. Reitzenstein, A. Knorr, A. Hoffmann,
  S. Kako, and Y. Arakawa, Phys. Rev. B {\bf 87},  245314  (2013).

\bibitem{Davanco2014}
M. Davan\c{c}o, C.~S. Hellberg, S. Ates, A. Badolato, and K. Srinivasan, Phys.
  Rev. B {\bf 89},  161303  (2014).

\bibitem{Wiersig2009}
J. Wiersig, C. Gies, F. Jahnke, M. Assmann, T. Berstermann, M. Bayer, C.
  Kistner, S. Reitzenstein, C. Schneider, S. H\"{o}fling, A. Forchel, C. Kruse,
  J. Kalden, and D. Hommel, Nature {\bf 460},  245  (2009).

\bibitem{Assmann2010a}
M. A\ss~mann, F. Veit, M. Bayer, C. Gies, F. Jahnke, S. Reitzenstein, S.
  H\"{o}fling, L. Worschech, and A. Forchel, Phys. Rev. B {\bf 81},  165314
  (2010).

\bibitem{Assmann2010b}
M. Assmann, F. Veit, J.-S. Tempel, T. Berstermann, H. Stolz, M. van~der Poel,
  J.~r.~M. Hvam, and M. Bayer, Opt. Express {\bf 18},  20229  (2010).

\bibitem{temnov09}
V.~V. Temnov and U. Woggon, Opt. Express {\bf 17},  5774  (2009).

\bibitem{auffeves11}
A. Auff\`{e}ves, D. Gerace, S. Portolan, A. Drezet, M. {Fran\c{c}a Santos}, and
  M.~F. Santos, New J. Phys. {\bf 13},  093020  (2011).

\bibitem{ritter10}
S. Ritter, P. Gartner, C. Gies, and F. Jahnke, Opt. Express {\bf 18},  9909
  (2010).

\bibitem{harsij12}
Z. Harsij, M. {Bagheri Harouni}, R. Roknizadeh, and M.~H. Naderi, Phys. Rev. A
  {\bf 86},  063803  (2012).

\bibitem{su10}
Y. Su, M. Richter, A. Knorr, D. Bimberg, and A. Carmele, Phys. status solidi -
  Rapid Res. Lett. {\bf 4},  289  (2010).

\bibitem{stephen64}
M.~J. Stephen, J. Chem. Phys. {\bf 40},  669  (1964).

\bibitem{lehmberg70a}
R.~H. Lehmberg, Phys. Rev. A {\bf 2},  883  (1970).

\bibitem{varfolomeev71}
A. Varfolomeev, Zh. Eksp. Teor. Fiz. {\bf 59},  1702  (1970).

\bibitem{Baumgartner2010}
A. Baumgartner, E. Stock, A. Patan\`{e}, L. Eaves, M. Henini, and D. Bimberg,
  Phys. Rev. Lett. {\bf 105},  257401  (2010).

\bibitem{Breuer}
H.-P. Breuer and F. Petruccione, {\em {The Theory of Open Quantum Systems}}
  (Oxford University Press, Oxford, 2007).

\bibitem{Gardiner2004}
C.~W. Gardiner and P. Zoller, {\em {Quantum noise: a handbook of Markovian and
  non-Markovian quantum stochastic methods with applications to quantum
  optics}} (Springer, Berlin, 2004).

\end{thebibliography}

\end{document}